# Electrolytic Synthesis and Characterizations of Silver Nanopowder

T. Theivasanthi* and M. Alagar

*Centre for Research and Post Graduate Department.of Physics, Ayya Nadar Janaki Ammal College, Sivakasi-626124, Tamilnadu, India.*

This work reports a simple, novel, cost effective and eco-friendly electrolytic synthesis of silver nanoparticles using $AgNO_3$ as metal precursor. The synthesis rate is much faster than other methods and this approach is suitable for large scale production. They are characterized by XRD, SEM and FT-IR techniques to analyze size, morphology and functional groups. XRD studies reveal a high degree of crystallinity and monophasic Ag nanoparticles. Their particle size is found to be 24 nm and specific surface area (SSA) is 24 $m^2/g$. Analysis of Ag nanoparticles SSA reports that increasing their SSA improves their antibacterial actions. Microbiology assay founds that Ag nanoparticles are effective against *E.coli* and *B.megaterium* bacteria. SSA of bacteria analysis reveals that it plays a major role while reacting with antimicrobial agents.

*Key Words:* XRD, Silver nanopowder, Williamson Hall plot, Electrolysis, Debye-Scherrer

## 1. Introduction

The synthesis of new materials made of particles, rods and wires with dimensions in the nanometer scale is among the most active areas of research in science due to the unique properties of these materials compared to conventional materials made from micron sized particles. In nanoparticles preparation, it is very important to control the particle size, particle shape and morphology. The characters of metal nanoparticles like optical, electronic, magnetic, and catalytic are depending on their size, shape and chemical surroundings [1].

New class of materials has been creating a technological revolution in the last decade. The common ground for these materials, and the devices made from them, is that they are constituted of building blocks of metals, ceramics or polymers that are nanometer size (1-100 nm) objects. The properties of such materials are novel and can be engineered by controlling the dimensions of these building blocks and their assembly via physical, chemical or biological methods. The basic physics and fundamental mechanisms responsible for nanoscale induced changes in properties will be stressed.

Most of the existing synthesizing methods of metal nanoparticles are complicated; require specific equipment and produce small amounts of nanomaterials. Metallic nanoparticles are traditionally synthesized by wet chemical methods where the chemicals used are often toxic and flammable. They are also synthesized by biosynthesis methods where there are a lot of difficulties to separate them and they are not ultra pure and stable only for few months.

Noble nanomaterials have been synthesized using a variety of methods, including hard template, bio-reduction and solution phase synthesis. Among noble metal nanomaterials, Silver nanoparticles have received considerable attention owing to their attractive physicochemical properties. The surface plasmon resonance and large effective scattering cross section of individual Silver nanoparticles make them ideal candidates for molecular labeling where phenomena such as surface enhanced Raman scattering (SERS) can be exploited. In addition, they exhibit strong toxicity in various chemical forms to a wide range of microorganism is well known and have shown to be a promising antimicrobial material.

---

*Corresponding author.    E-mail: theivasanthi@pacrpoly.org

Discussions about easy, simple, fast and low cost preparation i.e. Electrolysis synthesis of Silver nanopowder and its characterizations (XRD, SEM and FT-IR) are studied in this research paper. This method can be used to prepare wide range of materials. The synthesized nanoparticles size is 24nm. Also, in this paper we report changes in SSA of Silver nanoparticles and its effects on antibacterial activities and roles of bacteria SSA while reacting with Ag nanoparticles.

## 2. Experimental Details

Electrolysis method was adopted for Silver nanopowder preparation. 2gm of Silver Nitrate salt was kept in a cleaned glass vessel, 100 ml of distilled water was poured, stirred well and a homogenous aqueous Silver Nitrate solution was made. Surface-cleaned two Graphite rods are connected with Positive (Anode) & Negative (cathode) of a D.C.Power Supply unit (12 volt & 2 ampere) separately on one end, inserted in the Silver Nitrate solution on another end. Electrolysis of this solution was done by passing constant current inside solution through anode and cathode. At the end of electrolyzing process, Silver Nanoparticles deposition on the cathode surface was observed, they were removed carefully from the cathode surface and Silver nanopowder was collected. The size of the Nanoparticles was found to strongly depend on the concentration of solution & current. Its structural characterizations were studied and results confirmed the formation of Silver nanopowder.

XRD analysis of the prepared sample of Ag nanoparticles was done using a X'pert PRO of PANalytical diffractometer, Cu-K$\alpha$ X-rays of wavelength ($\lambda$)=1.54056 Å and data was taken for the 2$\theta$ range of 10° to 90° with a step of 0.02° The surface morphology were analyzed by using SEM Model S-3000H of HITACHI. Functional groups were analysed by SHIMADZU FT-IR spectrometer. The antibacterial activities of Ag nanoparticles were studied against *Escherichia coli* and *Bacillus megaterium* by Agar disc diffusion method. Standard zone of inhibition (ZOI) was measured and evaluated from this microbiology assay.

## 3. Results and Discussions

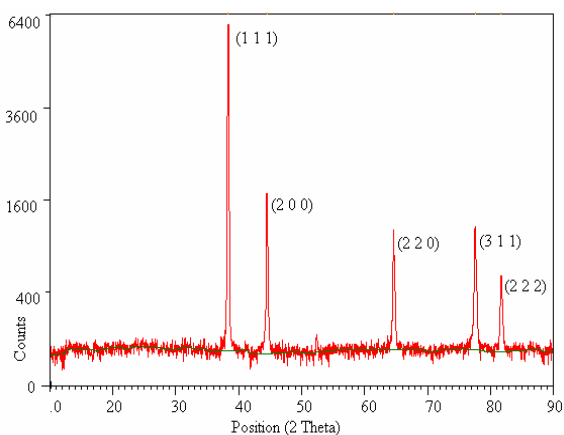
*Fig.1. XRD showing Peak Indices & 2 θ Positions*

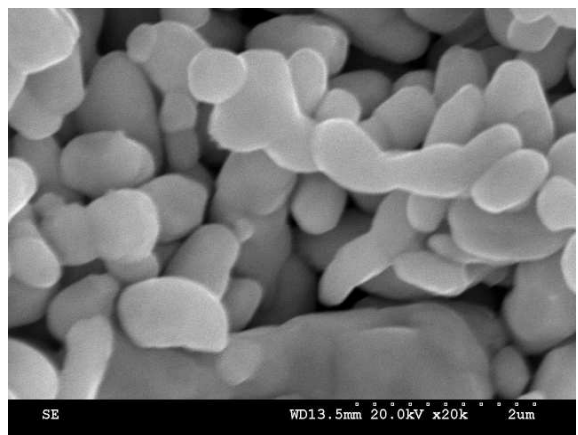
*Fig.2. SEM picture showing Silver Nanoparticles*

### 3.1. X-Ray Diffraction Studies - Peak Indexing

The X-ray diffraction pattern of the silver nanoparticles synthesized by electrolysis method is shown in Fig.1. Indexing process of powder diffraction pattern is done and *Miller Indices* (h k l) to each peak is assigned in first step. The details are in Table.1. A number of strong Bragg reflections can be seen which correspond to the (111), (200), (220) (311) and (222) reflections of fcc silver. No spurious diffractions due to crystallographic impurities are found [2]. Like this, no any spurious diffraction which indicating the crystallographic impurities in the sample. All the reflections correspond to pure silver metal with face centered cubic symmetry. The high intense peak for FCC materials is generally (1 1 1) reflection, which is observed in the sample. The intensity of peaks reflected the high degree of crystallinity of the silver nanoparticles. However, the diffraction peaks are broad which indicating that the crystallite size is very small [3]. The XRD shows that silver nanoparticles formed are crystalline. The size of the Ag nanoparticles estimated from the Debye–Scherrer formula (Instrumental broadening) is 24 nm.

*Table.1: Peak indexing from d – spacing*

| 2θ | d | 1000/$d^2$ | (1000/$d^2$) / 60.62 | hkl |
|---|---|---|---|---|
| 38.3182 | 2.35 | 181.10 | 3 | 111 |
| 44.4975 | 2.03 | 241.72 | 4 | 200 |
| 64.6119 | 1.44 | 482.30 | 8 | 220 |
| 77.5385 | 1.23 | 661.00 | 11 | 311 |
| 81.6839 | 1.18 | 718.20 | 12 | 222 |

Five peaks at 2θ values of 38.3182, 44.4975, 64.6119, 77.5385 and 81.6839 deg corresponding to (111), (200), (220), (311) and (222) planes of Silver is observed and compared with the standard powder diffraction card of Joint Committee on Powder Diffraction Standards (JCPDS), silver file No. 04–0783. The XRD study confirms / indicates that the resultant particles are (FCC) Silver Nanoparticles. [4]. The Experimental diffraction angle [2θ] and Standard diffraction angle [2θ] of the Table.2 are in agreement [5].

*Table.2: Experimental and standard diffraction angles of silver specimen*

| Experimental diffraction angle [2θ in degrees] | Standard diffraction angle [2θ in degrees] JCPDS Silver: 04-0783 |
|---|---|
| 44.4975 | 44.3 |

*Table.3: Ratio between the intensities of the diffraction peaks*

| Diffraction Peaks | Sample Value | Conventional Value |
|---|---|---|
| (200) and (111) | 0.26 | 0.40 |
| (220) and (111) | 0.16 | 0.25 |

The FCC crystal structure of silver has unit cell edge '*a*' = 4.07 Å and this value is calculated theoretically by using formula,

$$a = 4/\sqrt{2} \times r \quad \text{............(1)}$$

For silver r =144 pm. The experimental lattice constant '*a*' is calculated from the most intense peak (111) of the XRD pattern is 4.070 Å. Both theoretical & experimental lattice constant '*a*' are in very well agreement. The lattice constant '*a*' details have been produced in Table.4 and the values in agreement with the literature report (*a* = 4.086 Å, JCPDS file no. 04-0783). It is worth noting that the ratio between the intensities of the (200) & (111) diffraction peaks and (220) & (111) peaks enumerated in Table.3 is also slightly lower than the conventional value (0.26 versus 0.40) and (0.16 versus 0.25) [6].

### 3.2. XRD - Particle Size Calculation

From this study, considering the peak at degrees, average particle size has been estimated by using *Debye-Scherrer formula* [7, 8].

$$D = \frac{0.9\lambda}{\beta \cos\theta} \quad \quad \quad (2)$$

Where 'λ' is wave length of X-Ray (0.1541 nm), 'β' is FWHM (full width at half maximum), 'θ' is the diffraction angle and 'D' is particle diameter size. The calculated particle size details are in Table.4. The value of d (the interplanar spacing between the atoms) is calculated using *Bragg's Law* [9].

$$2d\sin\theta = n\lambda \quad \quad \quad (3)$$

*Table.4: The grain size of Silver nanopowder*

| 2θ of the intense peak (deg) | hkl | FWHM of Intense peak (β) radians | Size of the partcle (D) nm | d-spacing nm | Lattice parameter (a) Å |
|---|---|---|---|---|---|
| 38.3182 | (111) | 0.0041 | 36 | 0.235 | 4.070 |
| 44.4975 | (200) | 0.0048 | 31 | 0.203 | 4.068 |
| 64.6119 | (220) | 0.0038 | 43 | 0.144 | 4.073 |
| 77.5385 | (311) | 0.0034 | 52 | 0.123 | 4.079 |
| 81.6839 | (222) | 0.0038 | 49 | 0.118 | 4.088 |

### 3.3. XRD - Instrumental Broadening

When particle size is less than 100 nm, appreciable broadening in x-ray diffraction lines will occur. Diffraction pattern will show broadening because of particle size and strain. The observed line broadening will be used to estimate the average size of the particles. The total broadening of the diffraction peak is due to the sample and the instrument. The sample broadening is described by

$$FW(S) \times \cos\theta = \frac{K \times \lambda}{Size} + 4 \times Strain \times \sin\theta \quad \quad (4)$$

The total broadening $\beta_t$ is given by the equation

$$\beta_t^2 \approx \left\{\frac{0.9\lambda}{D\cos\theta}\right\}^2 + \{4\varepsilon \tan\theta\}^2 + \beta_0^2 \quad \quad (5)$$

$\varepsilon$ is strain and $\beta_0$ instrumental broadening. The average particle size D and the strain $\varepsilon$ of the experimentally observed broadening of several peaks will be computed simultaneously using *least squares method*. Instrumental Broadening is presented in Figure.3.

Williamson and Hall proposed a method for deconvoluting size and strain broadening by looking at the peak width as a function of 2θ. Here, Williamson-Hall plot is plotted with sin θ on the x-axis and $_\beta$ cos θ on the y-axis (in radians). A linear fit is got for the data. From this fit, particle size and strain are extracted from y-intercept and slope respectively [10]. The extracted particle size is 24 nm and strain is 0.0012. Figure.4. shows Williamson Hall Plot.

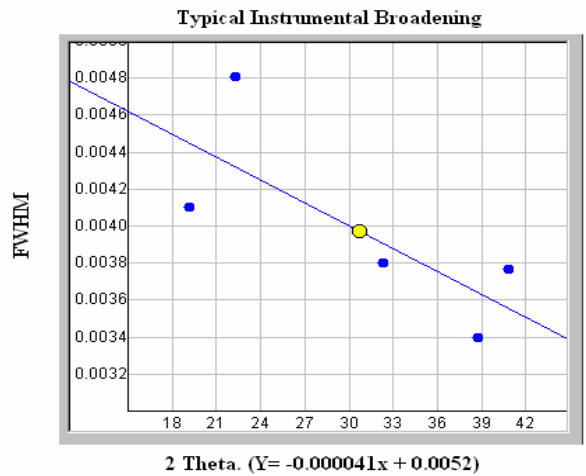
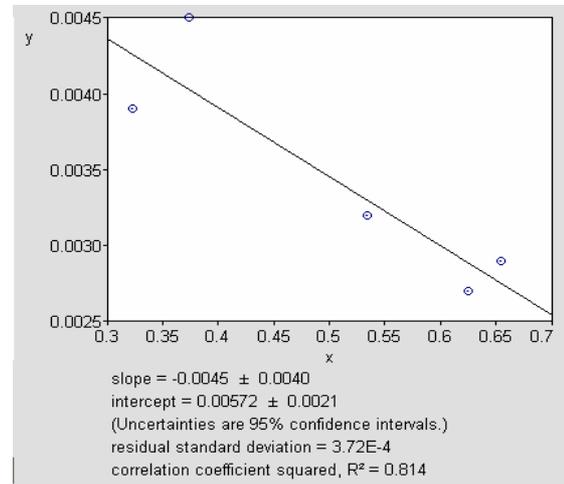

*Fig.3. Typical Instrumental Broadening.*
*y = -0.000041x + 0.0052*

*Fig.4. Williamson Hall Plot is indicating line broadening value due to the equipment.*

Line broadening analysis is most accurate when the broadening due to particle size effects is at least twice the contribution due to instrumental broadening. The size range is calculated over which this technique will be most accurate. A rough upper limit is estimated for reasonable accuracy by looking at the particle size that would lead to broadening equal to the instrumental broadening. For example, for Monochromatic Lab X-ray (Cu Kα FWHM ~ 0.05° at 20° 2θ), the accurate Size Range is < 90 nm (900 Å) and the rough Upper Limit is = < 180 nm (1800 Å).

### 3.4. XRD – Dislocation Density

In materials science, a dislocation is a crystallographic defect, or irregularity, within a crystal structure. The presence of dislocations strongly influences many of the properties of materials.  Mathematically, dislocations are a type of topological defect. The dislocation density increases with plastic deformation, a mechanism for the creation of dislocations must be activated in the material. Three mechanisms for dislocation formation are formed by homogeneous nucleation, grain boundary initiation, and interface the lattice and the surface, precipitates, dispersed phases, or reinforcing fibers.

The movement of a dislocation is impeded by other dislocations present in the sample. Thus, a larger *dislocation density* implies a larger hardness. Chen and Hendrickson measured and determined dislocation density and hardness of several silver crystals. They found that crystals with larger dislocation density were harder [11]. It has been shown for different pure face-centered cubic (fcc) metals processed by Equal Channel Angular Pressing (ECAP) that the dislocation density increases while the grain size decreases with increasing strain and ultimately these parameters reach saturation values [12].

It is well known that above a certain grain size limit (~20 nm) the strength of materials increases with decreasing grain size [13, 14]. The average dislocation density of silver is ~15 ± 2 ×$10^{14}$ m$^{-2}$ as obtained from the analysis of X-ray line profiles [15]. The X-ray line profile analysis has been used to determine the intrinsic stress and dislocation density of silver nanoparticles and found to be as 0.275 GPa and $7.0 \times 10^{14}$ m$^{-2}$ respectively [16].

The dislocation density (δ) in the sample has been determined using expression [17].

$$\delta = \frac{15\beta \cos\theta}{4aD} \quad \quad (6)$$

Where δ is dislocation density, $\beta$ is broadening of diffraction line measured at half of its maximum intensity (in radian), $\theta$ is Bragg's diffraction angle (in degree), $a$ is lattice constant (in nm) and $D$ is particle size (in nm). The dislocation density of sample silver nanoparticles found to be as $9.2 \times 10^{14}$ m$^{-2}$.

### 3.5. XRD - Crystallinity Index

It is generally agreed that the peak breadth of a specific phase of material is directly proportional to the mean crystallite size of that material. Quantitatively speaking, sharper XRD peaks are typically indicative of larger crystallite materials. From our XRD data, a peak broadening of the nanoparticles is noticed. The average particle size, as determined using the Scherrer equation, is calculated to be 24 nm. Crystallinity is evaluated through comparison of crystallite size as ascertained by SEM particle size determination. Crystallinity index Eq. is presented below:

$$I_{cry} = \frac{D_p(SEM, TEM)}{D_{cry}(XRD)} (I_{cry} \geq 1.00) \quad \quad (7)$$

Where $I_{cry}$ is the crystallinity index; $D_p$ is the particle size (obtained from either TEM or SEM morphological analysis); $D_{cry}$ is the particle size (calculated from the Scherrer equation).

*Table.5: The crystallinity index of Silver Nanoparticles*

| Sample | Dp (nm) | Dcry (nm) | Icry (unitless) | Particle Type |
|---|---|---|---|---|
| Silver Nanoparticles | 98 | 24 | ~4.08 | Polycrystalline |

Table.5. displays the crystallinity index of the sample that scored higher than 1.0. The data indicate that the silver metal is highly crystalline and fcc phase structure is well-indexed. If $I_{cry}$ value is close to 1, then it is assumed that the crystallite size represents monocrystalline whereas a polycrystalline have a much larger crystallinity index [18].

### 3.6. XRD - Specific Surface Area

Specific surface area (SSA) is a material property. It is a derived scientific value that can be used to determine the type and properties of a material. It has a particular importance in case of adsorption, heterogeneous catalysis and reactions on surfaces. SSA is the Surface Area (SA) per mass.

$$SSA = \frac{SA_{part}}{Vpart * density} \quad \quad (8)$$

Here Vpart is particle volume and SApart is particle SA [19].

$$S = 6 * 10^3 / D_p \rho \quad \text{................................} \quad (9)$$

Where S is the specific surface area, Dp is the size of the particles, and ρ is the density of silver 10.5 g/cm$^3$ [20]. Mathematically, SSA can be calculated using these formulas. Both of these formulas yield same result. Calculated value of SSA of the prepared silver nanoparticles is 24m$^2$/g.

### 3.7. XRD – Unit Cell Parameters

Unit cell parameters values calculated from XRD are enumerated in table.6.

*Table.6: XRD parameters of Silver Nanoparticles*

| Parameters | Values |
|---|---|
| Structure | FCC |
| Space group | Fm-3m (Space group number: 225) |
| Point group | m3m |
| Packing fraction | 0.74 |
| Symmetry of lattice | cubic close-packed |
| Particle size | 24 nm |
| Bond Angle | α = β = γ = 90˚ |
| Lattice parameters | a = b = c = 4.070 Å |
| Vol.unit cell(V) | 67.42 Å$^3$ |
| Radius of Atom | 144 pm |
| Density ( ρ ) | 10.5 g/cm$^3$ |
| Dislocation Density | 9.2 ×10$^{14}$ m$^{-2}$ |
| Mass | 107.8682 amu |

### 3.8. SEM morphological studies of Silver nanoparticles

Results of surface morphological and nanostructural studies using SEM are summarized in Fig. 2. The results indicate that mono-dispersive and highly crystalline Silver nanoparticles are obtained. The appearance is spherical in shape. The grain sizes of the samples estimated from the SEM picture is larger than that obtained from XRD data. This means that, the SEM picture indicates the size of polycrystalline particles. Generally, on the nanometer scale, metals (most of them are fcc) tend to nucleate and grow into twinned and multiply twinned particles (MTPs) with their surfaces bounded by the lowest-energy {111} facets. The observation of some larger nanoparticles may be attributed to the fact that Ag nanoparticles have the tendency to agglomerate due to their high surface energy and high surface tension of the ultrafine nanoparticles. The fine particle size results in a large surface area that in turn, enhances the nanoparticles catalytic activity.

### 3.9. FTIR Analyses of Silver nanoparticles:

FT-IR spectroscopic studies were carried out to investigate the plausible mechanism behind the formation of these silver nanoparticles and offer information regarding the functional groups. The representative spectra of Silver nanoparticles are shown in Fig. 5. Vibrational Assignments / Functional Groups corresponding to the absorption peaks are enumerated in Table.7. The very strong absorption peaks at 1624, 1600 and the strong absorption peaks at 1383, 1352 represents the presence of $NO_2$ which may be from $AgNO_3$ Solution, the metal precursor involved in the Ag nanoparticles synthesis process. Strong interaction of water with the surface of Silver could be the reason for the O-H stretching mode peaks at 2926, 2812, 2717 and O-H in plane bending mode peaks at 1383, 1352 [21, 22].

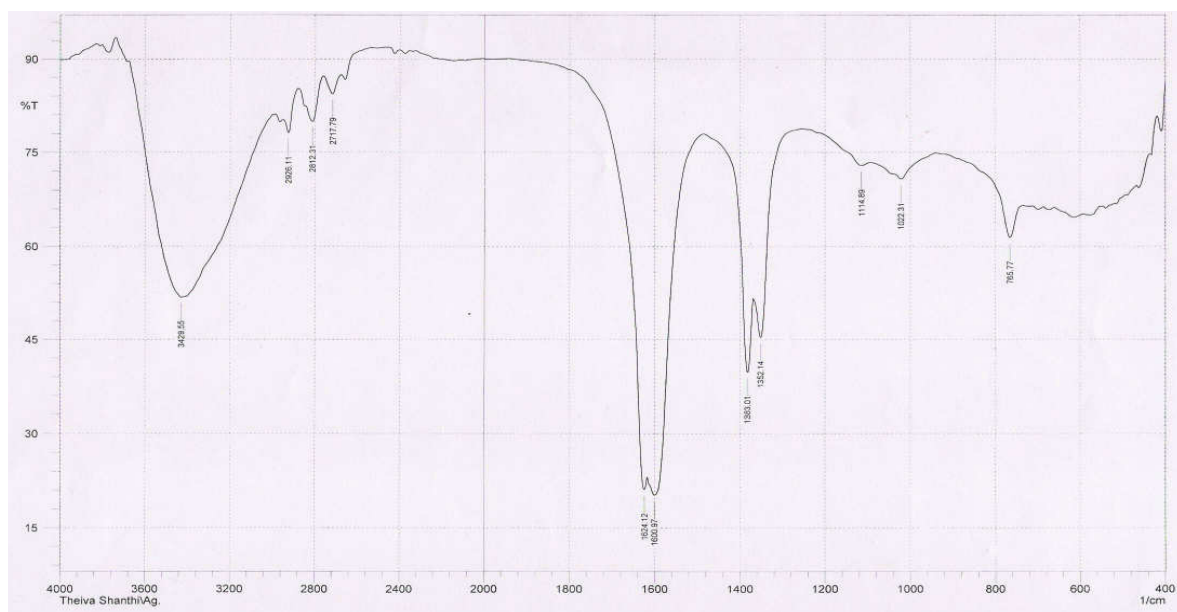

*Fig.5. Wave number ($cm^{-1}$) of dominant peaks obtained from FT-IR absorption spectra*

*Table.7: FTIR Functional groups analyses*

| Vibrational Assignment / Functional Groups | Observed Wave Number ($cm^{-1}$) | Visible Intensity |
|---|---|---|
| NH$\gamma_{as}$ | 3429.55 | S (Broad) |
| $CH_3$ $\gamma_{as}$ + $CH_2$ $\gamma_{as}$ + CH $\gamma$ + CH (Ketones) $\gamma$ + OH $\gamma$ | 2926.11 | W |
| OH $\gamma$ + $CH_3$ $\gamma$ + $CH_2$ $\gamma$ | 2812.31 | W |
| $CH_2$ $\gamma$ + CH $\gamma$ (Fermi resonance) + NH $\gamma$ + OH $\gamma$ | 2717.79 | W |
| NH $\alpha$ + $NO_2$ $\gamma_{as}$ + NO $\gamma$ | 1624.12 | VS |
| C=C $\gamma$ (vibration) + C$\cdots$C $\gamma$ (skeletal) + C=O $\gamma$ + COO$^-$ $\gamma_{as}$ + NH $\alpha$ + $NO_2$ $\gamma_{as}$ + $\delta_{as}$ $NH_3^+$ | 1600.97 | VS |
| $CH_3$ $\delta_s$ + $CH_2$ $\delta$ + OH $\alpha$ + C-O $\gamma$ + C-CHO (skeletal) + COO$^-$ $\gamma_s$ + $NO_2$ $\gamma_s$ | 1383.01 | S |
| $CH_3$ $\delta$ + OH $\alpha$ + C-O $\gamma$ + COO$^-$ $\gamma_s$ + C-N $\gamma$ + $NO_2$ $\gamma_s$ + $SO_2$ $\gamma_{as}$ + C=S $\gamma$ + C-F $\gamma$ | 1352.14 | S |
| C-O-C $\gamma_{as}$ + C-CHO (skeletal) + C-O $\gamma$ + N-N $\gamma$ + C-F $\gamma$ | 1114.89 | VW |
| Ring $\gamma$ + CH $\alpha$ + C-CHO (skeletal) + C-N $\gamma$ + N-N $\gamma$ + $SO_3$ $\gamma_s$ + S=O $\gamma$ + C-F $\gamma$ | 1022.31 | W |
| CH $\beta$ + C-O-C $\gamma$ + OCN (deformation) + N-H $\omega$ + N-H $\tau$ + O-N $\gamma$ + C-S $\gamma$ + C-Cl $\gamma$ | 765.77 | W |
| Abbreviations: $\alpha$ -in-plane bending; $\beta$ -out-of plane bending; $\omega$ –wagging; $\tau$ -twisting; $\gamma$ -stretching; $\delta$ – bending; $\gamma_s$ - symmetric stretching; $\gamma_{as}$ - asymmetric stretching; $\delta_s$ -symmetric bending; $\delta_{as}$ - asymmetric bending; | | |
| S – Strong; W – Weak; VS - Very strong; VW - Very Weak; | | |

## 3.10. Anti-Bacterial studies of Silver nanoparticles

Nanomaterials are the leading requirement of the rapidly developing field of nanomedicine, bionanotechnology. Nanoparticles usually have better or different qualities than the bulk material of the same element and have immense surface area relative to volume. For centuries, People have used silver for its antibacterial qualities. However, Silver Nanoparticles have showed antibacterial activities more than silver. Minuscule amounts of Silver Nanoparticles can lend antimicrobial effects to hundreds of square meters of its host material.

Antibacterial activities of silver nanoparticles synthesized by Electrolysis were evaluated by Agar disc diffusion method using Mueller hinton agar. Standard zone of inhibition (ZOI) was measured from this microbiology assay. The sample showed diameter of inhibition zone against *E.Coli* 12 mm & *B.megaterium* 6mm (Figures 6 and 7 show the results). In order to disclose the effective factors on their antibacterial activity, many studies have already been focused on the powder characteristics, such as particle size, shape and lattice constant by various researchers.

In a solid material, the surface-area-to-volume ratio (SA: V) or Specific Surface Area (SSA) is an important factor for the reactivity that is, the rate at which the chemical reaction will proceed. Materials with large SA: V (very small diameter) reacts at much faster rates than monolithic materials, because more surfaces are available to react.

For studying, changes in Specific Surface Area (SSA) of Silver Nanoparticles and its effects on antibacterial activities of Silver Nanoparticles, we have compared SSA of Silver Nanoparticles synthesized in Electrolysis method with extra cellular synthesis method of Silver Nanoparticles by Rajesh W. Raut et al. [23] and the details are in Table.8. From this comparative study, it is noted that antibacterial activities of Silver Nanoparticles prepared in Electrolysis method is more on *Escherichia Coli* (Gram Negative bacteria) than the Silver Nanoparticles prepared in Extra cellular synthesis method. It is also noted that increased SSA results in the enhancement of antibacterial activities of Silver nanoparticles.

*Table.8: Comparision of Surface Area to Volume Ratio and antibacterial activities of Silver Nanoparticles on Escherichia Coli (Gram Negative bacteria)*

| Silver Nanoparticles synthesis method | Particle Size (nm) | Surface Area (nm$^2$) | Volume (nm$^3$) | Surface Area to Volume Ratio | SSA (m$^2$/g) | Diameter Inhibition Zone (mm) |
|---|---|---|---|---|---|---|
| Electrolysis | 24 | 1809 | 7235 | 0.25 | 24 | 12 |
| Extra cellular synthesis | 38 | 4537 | 28735 | 0.18 | 15 | 8 |

The SSA of cells has an enormous impact on their biology. SSA places a maximum limit on the size of a cell. An increased SSA also means increased exposure to the environment. Greater SSA allows more of the surrounding water to be sifted for nutrients. Increased SSA can also lead to biological problems. More contact with the environment through the surface of a cell increases loss of water and dissolved substances. High SSA also present problems of temperature control in unfavorable environments.

SSA affects the rate at which particles can enter and exit the cell whereas the volume affects the rate at which material are made or used within the cell. These substances must diffuse between the organism and the surroundings. The rate at which a substance can diffuse is given by Fick's law.

$$\text{Rate of Diffusion} \propto \frac{\text{surface area} \times \text{concentration difference}}{\text{distance}} \quad \ldots\ldots (10)$$

So rate of exchange of substances depends on the organism's surface area that's in contact with the surroundings. Requirements for materials depend on the volume of the organism, so the ability to meet the requirements depends on the SSA.

*Table.9: Comparision of Bacteria Specific Surface Area and Silver Nanoparticles Actions*

| Name | Bacteria | | | Silver Nanoparticles | |
|---|---|---|---|---|---|
| | Variety | Specific Surface Area | Inhibition Zone Diameter | Specific Surface Area | Synthesis Method |
| *Escherichia coli* | Gram (-) | 20.09 m$^2$/g | 12 mm | 24 m$^2$/g | Electrolysis method |
| *Bacillus megaterium* | Gram (+) | 6.69 m$^2$/g | 6 mm | | |

In addition to Specific Surface Area (SSA) of Silver Nanoparticles, we have made an attempt to study the SSA of bacteria and its reactivity to antibacterial activities of Silver Nanoparticles. For this study, we have compared SSA of *E.coli* with *B.megaterium* and the details are in Table.9. *E.coli* details (Cell length: 2 μm or 2x10$^{-6}$ m, diameter: 0.8 μm or 0.8x10$^{-6}$ m, total volume: 1x10$^{-18}$ m$^3$, surface area: 6x10$^{-12}$ m$^2$, wet weight: 1x10$^{-12}$ g, dry weight: 3.0x10$^{-13}$ g) has been extracted from The CyberCell Database- CCDB and SSA calculated accordingly [24]. SSA of *B.megaterium* has been noted from the research of Rene Scherrer et al. [25]. From this analysis, we find that *E.coli* has more SSA than *B.megaterium*.

*Antibacterial activities evaluation of Electrolytic synthesized Silver nanoparticles (sample no.2)*

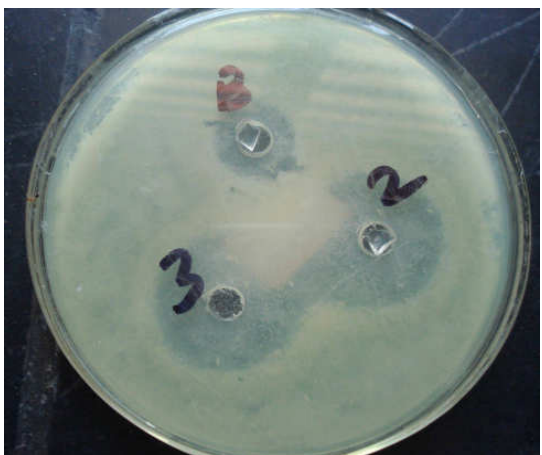 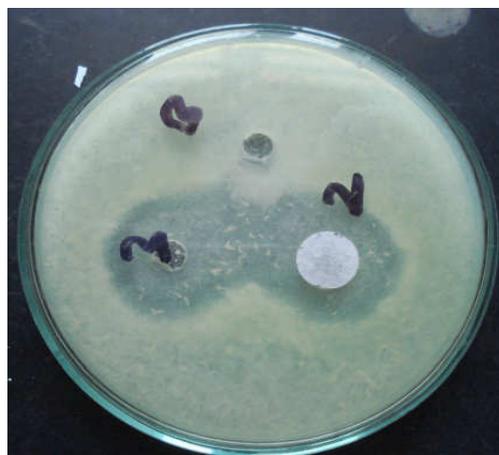

*Fig.6. Zone of inhibition diameter against Escherichia coli bacteria 12 mm*  *Fig.7. Zone of inhibition diameter against Bacillus megaterium 6mm*

More SSA of *E.coli* increases its exposure to the environment / surroundings in which Silver Nanoparticles are exist and this condition is unfavourable to *E.coli*. It increases rate of exchange of substances that is in contact with the surroundings of *E.coli*. It leads to more reactions of *E.coli* with Silver Nanoparticles than *B.megaterium*. Due to its more reactions in unfavourable surroundings results in increased Zone of Inhibition. Bacteria, viruses and fungi all depend on an enzyme to metabolize oxygen to live. Silver interferes with the effectiveness of the enzyme and disables the uptake of oxygen, thereby killing the microbes. This study reveals that the SSA of bacteria plays a major role while reacting with antimicrobial agents.

## 4. Conclusion

In conclusion, we introduce a simple, fast, and economical electrolysis method to synthesize Silver nanopowder. This method provides a clean, nontoxic and ecofriendly and efficient route for the synthesis of nanopowder with tunable particle size, at room temperature conditions without using any additive. There is no need to use high pressure, energy, temperature, toxic chemicals, downstream processing etc. Handling of the nanoparticles is also much easier than other methods. Based on this study, some other nanopowder may be prepared in future. From the point of view of nanotechnology, this is a significant advancement to synthesize silver nanopowder.

The synthesized Silver nanoparticles are in spherical shape with particle size of 24 nm. Their characterizations have been successfully done using XRD, SEM and FTIR spectroscopic techniques. Investigation on the antibacterial effect of nanosized silver against *E. coli* and *B.megaterium* microbes reveals high efficacy of silver nanoparticles as a strong antibacterial agent. SSA of Silver Nanoparticles prepared in two different methods have been analysed which concludes that increased SSA results in the enhancement of antibacterial activities of Silver nanoparticles. Likewise, analysis results of SSA of two different bacteria conclude that SSA of bacteria plays a major role while reacting with antimicrobial agents. This synthesized Silver nanopowder can be useful in food industries, cosmetic industries, medicines and other industries.

## Acknowledgements

The authors express immense thanks to **Dr.G.Venkadamanickam**, Rajiv Gandhi Cancer Institute & Research Center, Delhi, India, **Dr.M.Palanivelu**, *Arulmigu Kalasalingam College of Pharmacy (Kalasalingam University,* Krishnankoil, India), **S.Sivadevi,** *The SFR College for Women,* Sivakasi, India, staff & management of *PACR Polytechnic College*, Rajapalayam, India and *Ayya Nadar Janaki Ammal College*, Sivakasi, India for their valuable suggestions, assistances and encouragements during this work.